\documentclass[prb,twocolumn, superscriptaddress, floatfix]{revtex4}
\usepackage{graphicx}

\def\etal{{\em   et\ al.}}
\def\insitu{{\em in\ situ}}

\def\SRO{\mbox{Sr$_2$RuO$_{4}$}}

\begin{document}

\title{Cleaving-temperature dependence of layered-oxide surfaces}

\author{Y. Pennec}
\author{N.J.C. Ingle}
\email{ingle@physics.ubc.ca}
\author{I.S Elfimov}
\author{E. Varene}
\affiliation{Advanced Materials and Process Engineering Laboratory, University of British Columbia, Vancouver, BC, Canada}
\author{A. Damascelli}
\affiliation{Advanced Materials and Process Engineering Laboratory, University of British Columbia, Vancouver, BC, Canada}
\affiliation{Department of Physics and Astronomy, University of British Columbia, Vancouver, BC, Canada}
\author{J.V. Barth}
\affiliation{Advanced Materials and Process Engineering Laboratory, University of British Columbia, Vancouver, BC, Canada}
\affiliation{Department of Physics and Astronomy, University of British Columbia, Vancouver, BC, Canada}
\affiliation{Physik Department, TU M\"unchen, Munich, Germany}
\date{\today}
\author{Y. Maeno}
\affiliation{Department of Physics, Kyoto University, Kyoto 606-8502, Japan}

\begin{abstract}
The surfaces generated by cleaving non-polar, two-dimensional oxides are often considered to be perfect or ideal.  However, single particle spectroscopies on \SRO, an archetypal non-polar two dimensional oxide,  show significant cleavage temperature dependence.  We demonstrate that this is not a consequence of the intrinsic characteristics of the surface: lattice parameters and symmetries, step heights, atom positions, or density of states.   Instead, we find a marked increase in the density of defects at the mesoscopic scale with increased cleave temperature. The potential generality of these defects to oxide surfaces may have broad consequences to interfacial control and the interpretation of surface sensitive measurements. 

\end{abstract}
\maketitle

Cleaving, or fracturing, is often used as a means of generating a clean surface in vacuum, while ideally preserving the bulk stoichiometry and lateral structure in order to apply sophisticated surface sensitive measurement techniques such as angle resolved photoemission spectroscopy (ARPES) and scanning tunneling spectroscopy (STM).   This procedure has been applied to vast array of materials including semiconductors\cite{Bertelli07}, metals,\cite{Kolesnychenko01} and oxides.\cite{Lee06,Sun07,Moore07,Wakabayashi07}  In general, any material which does not readily allow the thermal regeneration of the bulk stoichiometry and structure from a polished, and hence mechanically damaged, surface is a candidate for using fracture as a means of generating a clean bulk representative surface \insitu.

\SRO\ is a prototypical system for using a cleaving procedure to generate a bulk representative surface.  It is also the first complex oxide providing quantitative agreement between bulk sensitive de Haas van Alphen (dHvA) measurements\cite{MacKenzie96,Bergemann03} and the surface sensitive ARPES measurements.\cite{Damascelli00, Ingle05, Shen07}  Structurally it is built up from a repeating three layer sequence of charge neutral SrO-RuO$_2$-SrO planes, which gives it a natural non-polar cleavage plane between neighboring SrO layers.  It also shows a large low temperature electrical anisotropy, $\frac{\rho_{ab}}{\rho_c}$, of $\sim4000$ indicating a strongly 2-dimensional electronic structure.\cite{MacKenzie03}  However, there is also clear evidence that the electronic structure, morphology, and even the presence of a superconducting gap on the surface depend on the exact cleavage parameters.\cite{Damascelli00,Upward02,Lupien05}  In fact, the ability to alter the electronic structure seen by ARPES by cleaving at different temperatures was the trick that allowed the initial quantitative connection between the aformentioned dHvA and ARPES measurements.  The high temperature cleaved surface shows a clear Fermi surface in ARPES that can be connected to the dHvA determined Fermi surface, while the low temperature cleave shows a much more complicated electronic structure that is now interpreted as resulting from a combination of the electronic structures of the bulk and of a reconstructed surface layer, first seen by Matzdorf \etal\cite{Matzdorf00}  This suggests that some level of stoichiometric or structural changes must occur as cleaving parameters change.  

In fact, from a fracture mechanics perspective, one would generally expect that temperature would play a large role on the exact formation of the cleavage plane. At it simplest, the role of temperature will reduce the energy needed to propagate a crack through the thermal activation of bond breaking. For example, the temperature dependence of the cleavage plane in B2 NiAl shows that decreasing temperature can cause the cleave plane to move away from the energetically more favorable surface.\cite{Ludwig98}  Temperature is also found to facilitate the introduction of dislocations in a metallic crystal which leads to a transition from brittle to ductile fracture.\cite{Cheung90}  Therefore, a variation of temperature may lead to very different final surfaces such as a different density of atomic step edges, a different cleaving plane, or a modified surface reconstruction.

In this letter we address the problem of the cleaving temperature dependence of the \SRO\ surface by systematically investigating the surface properties of samples cleaved at 20K and 200K, by means of STM.  The interpretation of the STM images is aided by Density Functional Theory (DFT) calculations.  We find that both high and low cleaving temperatures suggest a cleaving between the Sr-O planes in \SRO.  A closer look at both surfaces with atomic resolution shows the same very small modulation of the LDOS which has been connected with the rotation of the oxygen octahedral from the surface reconstruction seen in Low energy electron diffraction (LEED).\cite{Matzdorf00} However, we find a high number of randomly scattered point-like defects on the surface prepared at 200K.  These surface defects can effectively spread the spectral weight of quasiparticles originating from the reconstructed surface layer electronic structure over all k-space, thereby effectively removing it from the ARPES measurement and leaving only the bulk related electronic structure.  

All measurements were performed in an ultra-high vacuum chamber equipped with standard tools for surface preparation and characterization. The STM is a beetle-type operated at 8 K. We used W tips which were cleaned \insitu\ by Ar sputtering. A clean Au surface placed aside of the \SRO\ was used for tip sharpening through gentle contact between the tip and the Au surface. We studied a total of four samples coming from the same growth batch which had a T$_c$ of 1.26K indicating a mean free path of $\sim7000$\AA.\cite{MacKenzie98} A high and a low temperature cleave was performed \insitu, with identical vacuum conditions  200K and 20K, respectively. 

Although STM image currents are proportional to the matrix elements between the tip and the sample, they can be approximated by the local density of states (LDOS) at the Fermi energy of the surface as justified by Tersoff and Hamann.\cite{Tersoff85}    However, what is often shown from DFT calculations is the integration of the LDOS in an energy range around Ef, which is then formally a charge density.  In this paper we present the charge density contours as a function of lateral surface position to help interpret the STM images. DFT calculations were performed using the Siesta DFT package\cite{Soler02} with Troullier-Martins pseudo-potentials 
\cite{Troullier91} checked with the full-potential linearized augmented plane-wave DFT code WIEN2K\cite{Blaha01} by computing \SRO\ bulk properties. The exchange and correlation effects are treated within local density approximation after Ceperley and Alder.\cite{Ceperley80}  The slab geometry with 3 (defects) and 5 (no defects) formula-unit thickness was adopted throughout all surface calculations with vacuum region greater then 20\AA.  The 4x4 supercell in ab-plane was used to study a change in the electronic structure due to defects.

\begin{figure}
\includegraphics[width=0.47\textwidth]{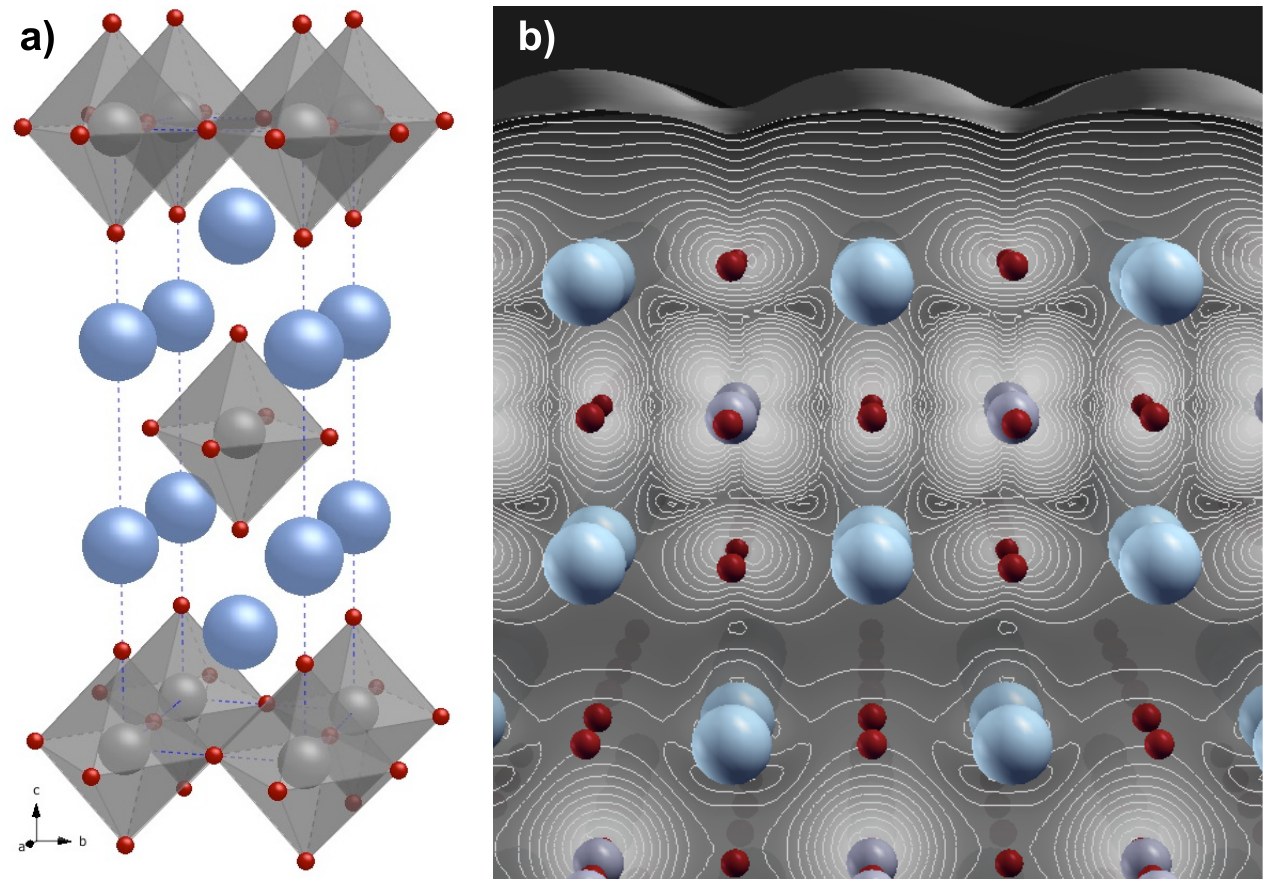}
\caption {(a) The unit cell of \SRO\ (Sr is blue, O is red,  and Ru is gray), showing the Ru centered RuO$_6$ octahedra.  (b) A model of \SRO\ with SrO (100) plane as the surface layer.  The iso-lines of the charge density, from the DFT calculations, are shown for the (100) plane through the Ru atoms at the rear of the image, and an iso-contour at a height of 2.13\AA\ above the surface is shown above the model. }
\label{siesta-surface-schematic}
\end{figure}

The first question arising when cleaving a complex crystallographic structure is what the most likely fracture plane is, and whether that plane could change at different cleaving temperatures.  In \SRO\ there are two possible cleavage plane candidates as shown by the arrows in Figure \ref{siesta-surface-schematic}a: between the SrO layers, or between the SrO-RuO$_2$ layers. 

STM work by Matzdorf \etal\cite{Matzdorf00} found extended terraces with consistent 6.4\AA\ step heights which they associated with the height of a complete Ru octahedral, suggesting that the cleavage plane is between the SrO-SrO layers in the (001) plane. If it is possible to cleave between the RuO$_2$-SrO planes one would expect to find at least two different step heights associated with either a single RuO$_2$ layer or a SrO-SrO-RuO$_2$ set.   Moreover, a recent surface specific chemical analysis of a very similar compound, La$_{0.5}$Sr$_{1.5}$MnO$_4$ found only a La/SrO terminated surface.\cite{Wakabayashi07}

Pseudo potential  calculations show that a surface generated by SrO-SrO planes is 0.141eV (per-surface and per-formula-unit) lower in energy than surfaces generated with a vacuum gap between the SrO-RuO2 planes in agreement with 0.138eV obtained from full-potential calculations.  This further supports the SrO termination of a cleaved surface.    Figure \ref{siesta-surface-schematic}b shows a slab of the SrO (001) terminated \SRO.  A charge density (integrating the LDOS from -100meV to $E_f$) isocontour, at a height of 2.13\AA\ above the SrO surface, is shown in the figure above the structure model.  At large distances above the surface ($> 1.7$\AA), which is typical of STM conditions, the height of the iso-contour is maximum directly above the Sr atom giving a square lateral modulation of spacing $3.87$\AA.  When looking at the iso-contours closer to the surface ($< 1.7$\AA), which would correspond to larger charge densities, it is found that the modulation changes form so as to have a maximum directly above O atoms in the SrO surface.  

\begin{figure}
\includegraphics[width=0.47\textwidth]{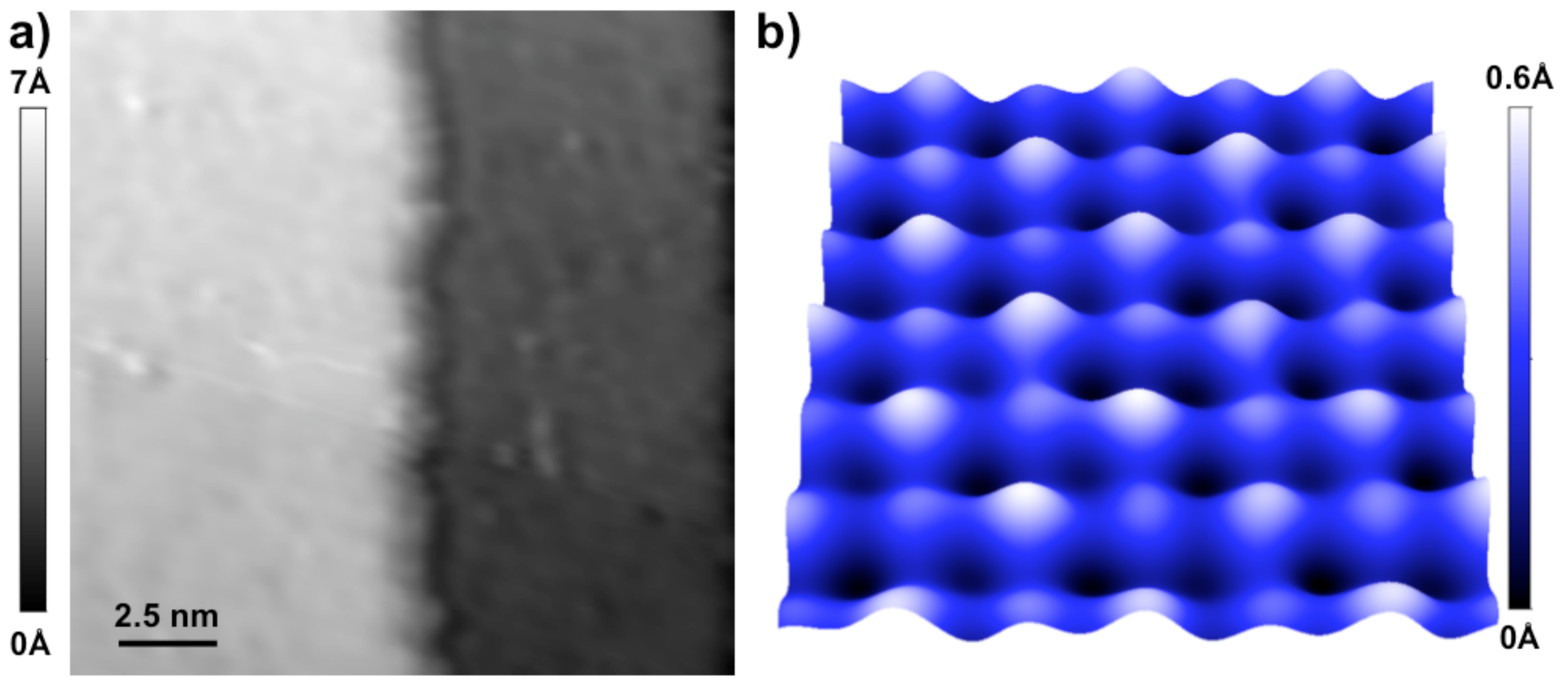}
\caption { Representative experimental STM images of the \SRO\ surface cleaved at both 20K and 200K.  (a) A $6.4 \pm 0.1$\AA\ high atomic step edge. (b) Atomic resolution within a terrace clearly showing the $\sqrt{2}\times\sqrt{2}$  surface reconstruction associated with the rotation of the Ru octahedra at the surface. }
\label{step-modulation-STM-image}
\end{figure}

For both high and low temperature cleaved samples, we experimentally find a few atomic step edges such as the one represented in Figure \ref{step-modulation-STM-image}a, usually one or less within our scanning range which is sub-micrometer. All of them correspond to the expected $6.4 \pm 0.1$\AA\ for a complete SrO-RuO$_2$-SrO unit.  We also find the distance between the maximum in the tunneling current has a lateral distance of $3.8\pm0.2$\AA.  Therefore we conclude that \SRO\ cleaves preferentially between the SrO planes regardless of the cleaving temperature, and that the maximum in the modulation we see in the STM image is occurring directly above the Sr atoms.  As a final note,  it has been found in an STM study of Sr$_2$(Ti$_{-x}$Ru$_{1-x}$)O$_4$ that the signature of the Ti substitution appears in between the STM maxima further supporting the idea of a corrugation centered on the Sr atoms.\cite{Barker03,Lupien05}

The different surface properties seen in ARPES\cite{Damascelli00, Shen01} and in STM\cite{Lupien05} could also be caused by different surface reconstructions. Using LEED, we found extra peaks associated with a $\sqrt{2}\times\sqrt{2}$  reconstruction (not shown), as discussed previously by Matzdorf \etal.\cite{Matzdorf02}.   This reconstruction is well known for \SRO\ and has been linked to a soft phonon mode inducing an $8.5^{\circ}$ rotation of the RuO$_6$ octahedra around the $c$-axis.\cite{Matzdorf02}  The LEED patterns from the two different temperature cleaves are indistinguishable.   Furthermore, from high resolution STM scans we find a similar topology for both temperatures cleaves.  As seen from a representative image in Figure \ref{step-modulation-STM-image}b, the atomic contrast in \SRO\ is weak, and the main corrugation, with the $3.8\pm0.2$\AA\ square lattice has an apparent tip height difference of 0.8\AA. In order to increase the signal to noise ratio we image the same area up to ten times and perform spatial averaging. Through this method we also find a secondary small modulation of intensity between two different Sr sites (up to an apparent 0.3\AA), as first seen by Matzdorf \etal\cite{Matzdorf00} This modulation is most likely the STM equivalent signature of the $\sqrt{2}\times\sqrt{2}$ reconstruction although we do not have any conclusive explanation for its presence in the STM experiment. Indeed, the two Sr sites on a reconstructed and relaxed surface are equivalent. Hence, there is no simple argument which would lead to either a height or LDOS difference between the two Sr sites. 

Finally we perform tunneling spectroscopy measurements on the two different temperature cleaves and also find the same signature for the electronic structure - which do not change if the measurement is performed on different Sr sites or on a O site. The tunneling spectra are V-shaped and non-zero at Ef which corresponds to a metallic surface. At this stage, we conclude that the cleaving temperature does not affect the fracture plane, the local surface topology, or the local surface electronic structure.

\begin{figure}
\includegraphics[width=0.47\textwidth]{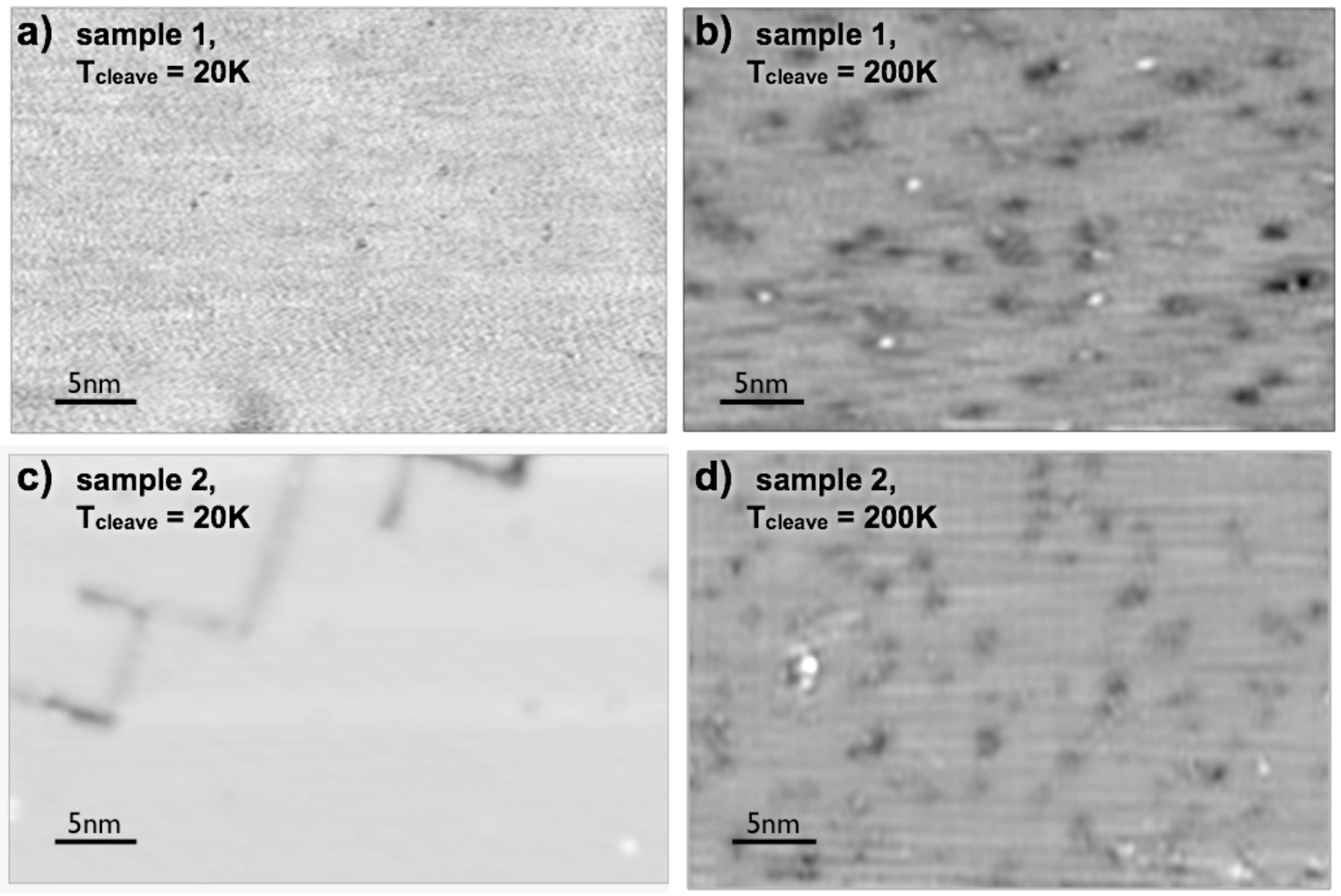}
\caption { STM topograph acquired from two different samples cleaved at both 20K, (a) and (c), and 200K, (b) and (d).  }
\label{defects-STM_image}
\end{figure}

To understand the cleavage temperature dependence seen in the ARPES spectra, one needs to look at an intermediate length scale inside each atomic terrace. Figure \ref{defects-STM_image} presents STM images acquired on two different samples, cleaved at both 20K (Fig. \ref{defects-STM_image}a,c) and at 200K (Fig. \ref{defects-STM_image}b,d). It appears clearly that the density of defects on the surface is directly related with the cleaving temperature. For the samples cleaved at 200K we find $0.056 \pm 0.01$ defects/nm$^2$, and the 20K cleaved sample shows $\sim0.002$ defects/nm$^2$. From the change in the density of defects between the two cleave temperatures and assuming a simple thermal activation of the defect formation, we can estimate the energy barrier for defect formation to be approximately 10 meV.  We also find that the 20K cleaved surfaces showed occasional highly ordered defects chains, propagating exactly $45^\circ$ off the main crystallographic axis and following the average direction of the cleave, as seen in the upper left corner of Figure \ref{defects-STM_image},c.

While the surface reconstruction and cleavage plane are identical in both low- and high-temperature cleaves of \SRO, the defect concentrations are very different for the two cleave temperatures. It is thus tempting to connect the selective suppression of the reconstructed-surface layer electronic structure observed in ARPES on the high-temperature cleaves,\cite{Damascelli00} to the higher uniform density of point-like defects revealed by our STM study. In particular, this suppression would have to be a consequence of the increased impurity scattering for the electrons in the top-most and reconstructed RuO$_2$ plane, due to the defects introduced by the cleaving process in the SrO termination plane right above. We note that the close proximity between the SrO termination plane and the first RuO$_2$ plane implies much stronger scattering effects for the electrons propagating in first RuO$_2$ plane than in the deeper ones. Also, because the surface electronic structure is characterized by very flat bands with extended van Hove singularities a few meV above and below the Fermi energy,\cite{Shen07} elastic impurity scattering would lead to a strong suppression of momentum-resolved ARPES features for the surface than it would for the bulk; this suppression is accompanied by an increase of the angle-independent background.

\begin{figure}
\includegraphics[width=0.49\textwidth]{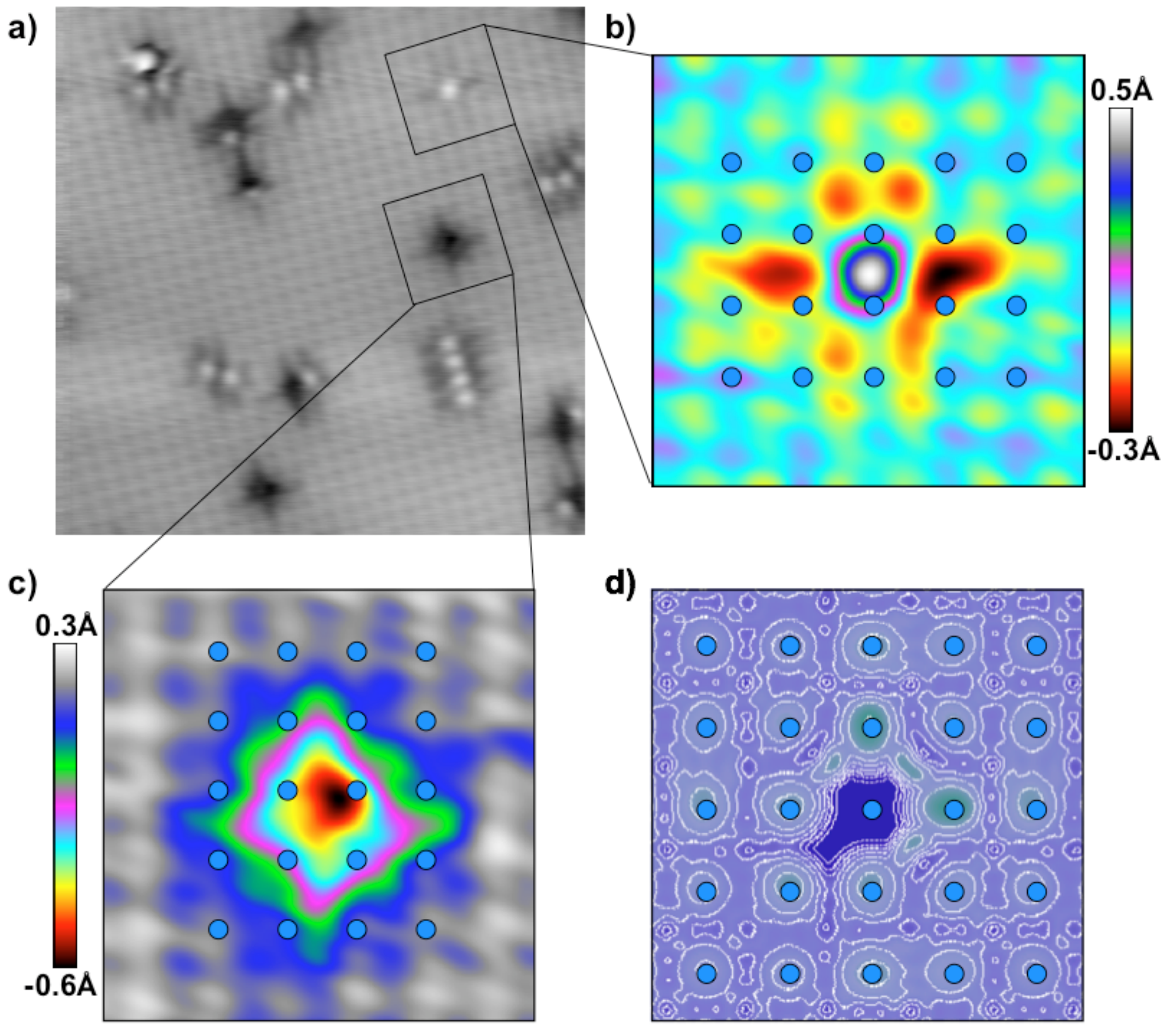}
\caption { (a) $10\times10$ nm$^2$ STM images with atomic resolution of the \SRO\ surface obtained from a 200K cleave. The two types of characteristic defects are show with false color maps in (b), the {\em  protrusion}, and (c), the {\em hole}.  Blue dots are the Sr locations on the SrO terminated surface.   (d) The fully relaxed DFT calculated iso-contour of the charge density at a height of 2.13\AA, for a charge neutral SrO molecule missing from the surface.  }
\label{defects-zoom-STM_image}
\end{figure}

A high resolution STM image of the 200K cleaved sample (Figure \ref{defects-zoom-STM_image}) shows clusters of defects developing along the crystallographic axis of the surface. The clusters are built on the juxtaposition of two primal objects: a {\em hole} and a {\em protrusion}. This duality would seem to suggest that the hole and the protrusion could be two matching pieces of a jigsaw puzzle: during the cleaving process an atom, or molecule, normally belonging to the bottom surface sticks instead to the upper surface.   

The two close up STM images of Figure \ref{defects-zoom-STM_image}b and c, with artificial color map for an increased contrast, provide a better insight on the nature of the defects -- note that those two images come from a single acquired STM in order to avoid any possible image treatment artifact.  The {\em hole} defect object is shown in Figure \ref{defects-zoom-STM_image}c.  The core of the hole is located at a Sr site but with a clear left/bottom lateral shift towards the O.  The registry of this site with the SrO termination plane suggest this defect is a missing atom or molecule. In Figure \ref{defects-zoom-STM_image}b, the {\em protrusion} defect site is quite symmetric and sits half way between to Sr sites.  This location is not in registry with any atoms in the SrO termination plane, but is above one of the O atoms of the deeper RuO$_2$ plane.  This location suggests the defect sits above the SrO termination plane.

One intriguing question is whether these defects are molecular in nature or atomic-like and how generic they may be to two dimensional oxides.  DFT calculations of fully relaxed SrO terminated surfaces, which include a missing Sr atom, O atom, or SrO molecule, give cohesive energies for all defects of approximately $0.2 \pm 0.03$eV, with energy differences between the different defects that are smaller than the thermal energy at room temperature ($\sim 25$ meV).  It is also found that thermal evaporation of SrO leads to both Sr atoms and SrO molecules in the gas phase.\cite{Moore50}  These facts do not point to a particular preference for the type of defect expected.   However, the DFT calculations do suggest that a missing charge neutral SrO is the only defect that generates a charge density contour that is topologically similar to the {\em hole}.  The missing Sr and O defects both generate a highly symmetric topology centered directly over the respective lattice site.  

This work, which demonstrates the temperature dependent cleave generation of surface defects, provides a natural explanation of the loss of the reconstructed surface-layer electronic structure in ARPES while leaving the surface reconstruction visible in LEED.   The defects show two distinct topologies, a {\em protrusion} and a {\em hole} and DFT calculations suggest the {\em hole} defect is a missing charge neutral SrO molecule.   The effect of these potentially generic cleave generated defects are uniquely distinguishable by ARPES in \SRO\ because the presence of the surface layer reconstruction generates an altered electronic structure of the surface which is clearly distinguishable from the bulk electronic structure.   This, in essence, allows a depth dependent measurement of the scattering effects of a defect in the "insulating" surface layer of two dimensional oxides by surface sensitive measurements.


\begin{thebibliography}{26}
\expandafter\ifx\csname natexlab\endcsname\relax\def\natexlab#1{#1}\fi
\expandafter\ifx\csname bibnamefont\endcsname\relax
  \def\bibnamefont#1{#1}\fi
\expandafter\ifx\csname bibfnamefont\endcsname\relax
  \def\bibfnamefont#1{#1}\fi
\expandafter\ifx\csname citenamefont\endcsname\relax
  \def\citenamefont#1{#1}\fi
\expandafter\ifx\csname url\endcsname\relax
  \def\url#1{\texttt{#1}}\fi
\expandafter\ifx\csname urlprefix\endcsname\relax\def\urlprefix{URL }\fi
\providecommand{\bibinfo}[2]{#2}
\providecommand{\eprint}[2][]{\url{#2}}

\bibitem[{\citenamefont{Bertelli et~al.}(2007)\citenamefont{Bertelli, Homoth,
  Wenderoth, Rizzi, Ulbrich, Righi, Bertoni, and Catellani}}]{Bertelli07}
\bibinfo{author}{\bibfnamefont{M.}~\bibnamefont{Bertelli}},
  \bibnamefont{et~al.},  \bibinfo{journal}{Phys. Rev. B} \textbf{\bibinfo{volume}{75}},
  \bibinfo{pages}{165312} (\bibinfo{year}{2007}).

\bibitem[{\citenamefont{Kolesnychenko et~al.}(2001)\citenamefont{Kolesnychenko,
  de~Kort, and van Kempen}}]{Kolesnychenko01}
\bibinfo{author}{\bibfnamefont{O.}~\bibnamefont{Kolesnychenko}},
  \bibinfo{author}{\bibfnamefont{R.}~\bibnamefont{de~Kort}}, \bibnamefont{and}
  \bibinfo{author}{\bibfnamefont{H.}~\bibnamefont{van Kempen}},
  \bibinfo{journal}{Surface Science} \textbf{\bibinfo{volume}{490}},
  \bibinfo{pages}{L573} (\bibinfo{year}{2001}).

\bibitem[{\citenamefont{Lee et~al.}(2006)\citenamefont{Lee, Fujita, McElroy,
  Slezak, Wang, Aiura, Bando, Ishikado, Masui, Zhu et~al.}}]{Lee06}
\bibinfo{author}{\bibfnamefont{J.}~\bibnamefont{Lee}},
    \bibnamefont{et~al.},\bibnamefont{et~al.}, \bibinfo{journal}{Nature}
  \textbf{\bibinfo{volume}{442}}, \bibinfo{pages}{546} (\bibinfo{year}{2006}).

\bibitem[{\citenamefont{Sun et~al.}(2007)\citenamefont{Sun, Douglas, Fedorov,
  Chuang, Zheng, Mitchell, and Dessau}}]{Sun07}
\bibinfo{author}{\bibfnamefont{Z.}~\bibnamefont{Sun}},
 \bibnamefont{et~al.},\bibinfo{journal}{Nature Physics}
  \textbf{\bibinfo{volume}{3}}, \bibinfo{pages}{248} (\bibinfo{year}{2007}).

\bibitem[{\citenamefont{Moore et~al.}(2007)\citenamefont{Moore, Zhang,
  Nascimento, Jin, Guo, Wang, Fang, Mandrus, and Plummer}}]{Moore07}
\bibinfo{author}{\bibfnamefont{R.~G.} \bibnamefont{Moore}},
  \bibnamefont{et~al.},  \bibinfo{journal}{Science} \textbf{\bibinfo{volume}{318}},
  \bibinfo{pages}{615} (\bibinfo{year}{2007}).

\bibitem[{\citenamefont{Wakabayashi et~al.}(2007)\citenamefont{Wakabayashi,
  Upton, Grenier, Hill, Nelson, Kim, Ryan, Goldman, Zheng, and
  Mitchell}}]{Wakabayashi07}
\bibinfo{author}{\bibfnamefont{Y.}~\bibnamefont{Wakabayashi}},
   \bibnamefont{et~al.}, \bibinfo{journal}{Nature Materials} \textbf{\bibinfo{volume}{6}},
  \bibinfo{pages}{972} (\bibinfo{year}{2007}).

\bibitem[{\citenamefont{Bergemann et~al.}(2003)\citenamefont{Bergemann,
  Mackenzie, Julian, Forsythe, and Ohmichi}}]{Bergemann03}
\bibinfo{author}{\bibfnamefont{C.}~\bibnamefont{Bergemann}},
   \bibnamefont{et~al.}, \bibinfo{journal}{Advances In Physics} \textbf{\bibinfo{volume}{52}},
  \bibinfo{pages}{639} (\bibinfo{year}{2003}).

\bibitem[{\citenamefont{Mackenzie et~al.}(1996)\citenamefont{Mackenzie, Julian,
  Diver, McMullan, Ray, Lonzarich, Maeno, Nishizaki, and Fujita}}]{MacKenzie96}
\bibinfo{author}{\bibfnamefont{A.}~\bibnamefont{Mackenzie}},
   \bibnamefont{et~al.}, \bibinfo{journal}{Phys. Rev. Let.} \textbf{\bibinfo{volume}{76}},
  \bibinfo{pages}{3786} (\bibinfo{year}{1996}).

\bibitem[{\citenamefont{Damascelli et~al.}(2000)\citenamefont{Damascelli, Lu,
  Shen, Armitage, Ronning, Feng, Kim, Shen, Kimura, Tokura
  et~al.}}]{Damascelli00}
\bibinfo{author}{\bibfnamefont{A.}~\bibnamefont{Damascelli}},
  \bibnamefont{et~al.}, \bibinfo{journal}{Phys. Rev. Let.}
  \textbf{\bibinfo{volume}{85}}, \bibinfo{pages}{5194} (\bibinfo{year}{2000}).

\bibitem[{\citenamefont{Ingle et~al.}(2005)\citenamefont{Ingle, Shen,
  Baumberger, Meevasana, Lu, Shen, Damascelli, Nakatsuji, Mao, Maeno
  et~al.}}]{Ingle05}
\bibinfo{author}{\bibfnamefont{N.}~\bibnamefont{Ingle}},
  \bibnamefont{et~al.}, \bibinfo{journal}{Phys. Rev. B}
  \textbf{\bibinfo{volume}{72}}, \bibinfo{pages}{205114}
  (\bibinfo{year}{2005}).

\bibitem[{\citenamefont{Shen et~al.}(2007)\citenamefont{Shen, Kikugawa,
  Bergemann, Balicas, Baumberger, Meevasana, Ingle, Maeno, Shen, and
  Mackenzie}}]{Shen07}
\bibinfo{author}{\bibfnamefont{K.~M.} \bibnamefont{Shen}},
  \bibnamefont{et~al.},  \bibinfo{journal}{Phys. Rev. Let.} \textbf{\bibinfo{volume}{99}},
  \bibinfo{pages}{187001} (\bibinfo{year}{2007}).

\bibitem[{\citenamefont{Mackenzie and Maeno}(2003)}]{MacKenzie03}
\bibinfo{author}{\bibfnamefont{A.}~\bibnamefont{Mackenzie}} \bibnamefont{and}
  \bibinfo{author}{\bibfnamefont{Y.}~\bibnamefont{Maeno}},
  \bibinfo{journal}{Reviews of Modern Physics} \textbf{\bibinfo{volume}{75}},
  \bibinfo{pages}{657} (\bibinfo{year}{2003}).

\bibitem[{\citenamefont{Upward et~al.}(2002)\citenamefont{Upward, Kouwenhoven,
  Morpurgo, Kikugawa, Mao, and Maeno}}]{Upward02}
\bibinfo{author}{\bibfnamefont{M.}~\bibnamefont{Upward}},
   \bibnamefont{et~al.}, \bibinfo{journal}{Phys. Rev. B} \textbf{\bibinfo{volume}{65}},
  \bibinfo{pages}{220512} (\bibinfo{year}{2002}).

\bibitem[{\citenamefont{Lupien et~al.}(2005)\citenamefont{Lupien, Dutta, and
  Y.~Maeno, and Davis}}]{Lupien05}
\bibinfo{author}{\bibfnamefont{C.}~\bibnamefont{Lupien}},
   \bibnamefont{et~al.}, \bibinfo{journal}{arXiv:cond-mat/0503317}  (\bibinfo{year}{2005}).

\bibitem[{\citenamefont{Shen et~al.}(2001)\citenamefont{Shen, Damascelli, Lu,
  Armitage, Ronning, Feng, Kim, Shen, Singh, Mazin et~al.}}]{Shen01}
\bibinfo{author}{\bibfnamefont{K.}~\bibnamefont{Shen}},
  \bibnamefont{et~al.},  \bibnamefont{et~al.}, \bibinfo{journal}{Phys. Rev. B}
  \textbf{\bibinfo{volume}{64}}, \bibinfo{pages}{180502}
  (\bibinfo{year}{2001}).

\bibitem[{\citenamefont{Matzdorf et~al.}(2000)\citenamefont{Matzdorf, Fang,
  Ismail, Zhang, Kimura, Tokura, Terakura, and Plummer}}]{Matzdorf00}
\bibinfo{author}{\bibfnamefont{R.}~\bibnamefont{Matzdorf}},
   \bibnamefont{et~al.}, \bibinfo{journal}{Science} \textbf{\bibinfo{volume}{289}},
  \bibinfo{pages}{746} (\bibinfo{year}{2000}).

\bibitem[{\citenamefont{Ludwig and Gumbsch}(1998)}]{Ludwig98}
\bibinfo{author}{\bibfnamefont{M.}~\bibnamefont{Ludwig}} \bibnamefont{and}
  \bibinfo{author}{\bibfnamefont{P.}~\bibnamefont{Gumbsch}},
  \bibinfo{journal}{Acta Materialia} \textbf{\bibinfo{volume}{46}},
  \bibinfo{pages}{3135} (\bibinfo{year}{1998}).

\bibitem[{\citenamefont{Cheung and Yip}(1990)}]{Cheung90}
\bibinfo{author}{\bibfnamefont{K.}~\bibnamefont{Cheung}} \bibnamefont{and}
  \bibinfo{author}{\bibfnamefont{S.}~\bibnamefont{Yip}},
  \bibinfo{journal}{Physical Review Letters} \textbf{\bibinfo{volume}{65}},
  \bibinfo{pages}{2804} (\bibinfo{year}{1990}).

\bibitem[{\citenamefont{Mackenzie et~al.}(1998)\citenamefont{Mackenzie,
  Haselwimmer, Tyler, Lonzarich, Mori, Nishizaki, and Maeno}}]{MacKenzie98}
\bibinfo{author}{\bibfnamefont{A.}~\bibnamefont{Mackenzie}},
  \bibnamefont{et~al.},  \bibinfo{journal}{Phys. Rev. Let.} \textbf{\bibinfo{volume}{80}},
  \bibinfo{pages}{3890} (\bibinfo{year}{1998}).

\bibitem[{\citenamefont{Tersoff and Hamann}(1985)}]{Tersoff85}
\bibinfo{author}{\bibfnamefont{J.}~\bibnamefont{Tersoff}} \bibnamefont{and}
  \bibinfo{author}{\bibfnamefont{D.}~\bibnamefont{Hamann}},
  \bibinfo{journal}{Phys. Rev. B} \textbf{\bibinfo{volume}{31}},
  \bibinfo{pages}{805} (\bibinfo{year}{1985}).

\bibitem[{\citenamefont{Soler et~al.}(2002)\citenamefont{Soler, Artacho, Gale,
  Garcia, Junquera, Ordejon, and Sanchez-Portal}}]{Soler02}
\bibinfo{author}{\bibfnamefont{J.}~\bibnamefont{Soler}},
   \bibnamefont{et~al.}, \bibinfo{journal}{Journal of Physics-Condensed Matter}
  \textbf{\bibinfo{volume}{14}}, \bibinfo{pages}{2745} (\bibinfo{year}{2002}).

\bibitem[{\citenamefont{Troullier and Martins}(1991)}]{Troullier91}
\bibinfo{author}{\bibfnamefont{N.}~\bibnamefont{Troullier}} \bibnamefont{and}
  \bibinfo{author}{\bibfnamefont{J.}~\bibnamefont{Martins}},
  \bibinfo{journal}{Phys. Rev. B} \textbf{\bibinfo{volume}{43}},
  \bibinfo{pages}{8861} (\bibinfo{year}{1991}).

\bibitem[{\citenamefont{Blaha et~al.}(2001)\citenamefont{Blaha, Schwarz,
  Madsen, Kvasnicka, and Luitz}}]{Blaha01}
\bibinfo{author}{\bibfnamefont{P.}~\bibnamefont{Blaha}},
  \bibnamefont{et~al.},  \emph{\bibinfo{title}{{WIEN2k}: An augmented plane wave plus local orbitals
  program for calculating crystal properties}} (\bibinfo{publisher}{TU Wien,
  Austria}, \bibinfo{year}{2001}).

\bibitem[{\citenamefont{Ceperley and Alder}(1980)}]{Ceperley80}
\bibinfo{author}{\bibfnamefont{D.}~\bibnamefont{Ceperley}} \bibnamefont{and}
  \bibinfo{author}{\bibfnamefont{B.}~\bibnamefont{Alder}},
  \bibinfo{journal}{Phys. Rev. Let.} \textbf{\bibinfo{volume}{45}},
  \bibinfo{pages}{566} (\bibinfo{year}{1980}).

\bibitem[{\citenamefont{Barker et~al.}(2003)\citenamefont{Barker, Dutta,
  Lupien, McEuen, Kikugawa, Maeno, and Davis}}]{Barker03}
\bibinfo{author}{\bibfnamefont{B.}~\bibnamefont{Barker}},
   \bibnamefont{et~al.}, \bibinfo{journal}{Physica B-Condensed Matter} \textbf{\bibinfo{volume}{329}},
  \bibinfo{pages}{1334} (\bibinfo{year}{2003}).

\bibitem[{\citenamefont{Matzdorf et~al.}(2002)\citenamefont{Matzdorf, Ismail,
  Kimura, Tokura, and Plummer}}]{Matzdorf02}
\bibinfo{author}{\bibfnamefont{R.}~\bibnamefont{Matzdorf}},
   \bibnamefont{et~al.}, \bibinfo{journal}{Phys. Rev. B} \textbf{\bibinfo{volume}{65}},
  \bibinfo{pages}{085404} (\bibinfo{year}{2002}).

\bibitem[{\citenamefont{Moore and Allison}(1950)}]{Moore50}
\bibinfo{author}{\bibfnamefont{G.}~\bibnamefont{Moore}} \bibnamefont{and}
  \bibinfo{author}{\bibfnamefont{H.}~\bibnamefont{Allison}},
  \bibinfo{journal}{Phyiscal Review} \textbf{\bibinfo{volume}{77}},
  \bibinfo{pages}{246} (\bibinfo{year}{1950}).

\end{thebibliography}
\end{document}